\documentclass[11pt,a4paper]{article}
\usepackage{authblk}
\usepackage[margin=1in]{geometry}
\usepackage[hidelinks]{hyperref}
\usepackage[utf8]{inputenc}
\usepackage[T1]{fontenc}
\usepackage{amsmath,amssymb,amsfonts}
\usepackage{graphicx}
\usepackage{booktabs}
\usepackage{tikz}
\usepackage{pgfplots}
\pgfplotsset{compat=1.18}
\usepackage{siunitx}
\usetikzlibrary{arrows.meta}

\begin{document}

\title{Cognitive Amplification vs Cognitive Delegation in Human-AI Systems: A Metric Framework}

\author[1]{Eduardo Di Santi\thanks{eduardo.disanti@colorado.edu}}
\author[2]{Carla Florida\thanks{carla.florida.mpsia@est.umss.edu}}

\affil[1]{College of Engineering and Applied Science, University of Colorado Boulder}
\affil[2]{Facultad de Ciencias y Tecnología, Universidad Mayor de San Simon}

\date{April 2026}

\maketitle

\begin{abstract}
Artificial intelligence is increasingly embedded in human decision-making, yet the distinction between systems that genuinely amplify human cognition and those that promote excessive dependence remains operationally underdefined. This paper introduces a compact evaluative framework for distinguishing cognitive amplification---where hybrid performance improves without degrading human capability---from cognitive delegation, where reasoning is progressively outsourced to the system.

We define four operational metrics: the Cognitive Amplification Index ($CAI^*$), the Dependency Ratio ($D$), the Human Reliance Index ($HRI$), and the Human Cognitive Drift Rate ($HCDR$). We test the structural discriminative capacity of this framework in a stylized agent-based simulation in NetLogo across three reliance regimes and multiple dependency--atrophy configurations. We then perform a constrained optimization over the atrophy parameter and additional parameter sweeps to determine whether positive collaborative gain can be recovered within the baseline model. Finally, we introduce a structural extension with an explicit human--AI interaction term.

Within the baseline simulated interaction regimes, the proposed metrics distinguish degenerate AI-dominated delegation, interaction that is more capability-preserving than full delegation but weakly competitive, and structurally dependent boundary regimes. Across all baseline configurations, no regime achieves positive collaborative gain relative to the best standalone baseline, and reducing atrophy to zero remains insufficient. Additional sweeps show that this limitation is structural rather than merely parametric. Positive $CAI^*>0$ becomes attainable only after introducing an explicit interaction term that allows retained human capability to contribute to the assisted output.

The proposed framework provides a practical basis for evaluating not only whether human--AI systems perform well, but whether they remain cognitively sustainable over time. More broadly, the results suggest that preventing capability erosion alone is insufficient for genuine amplification when the interaction architecture remains delegation-oriented. Amplification requires both preserved human capability and a coupling mechanism through which that capability contributes productively to the hybrid outcome.

\vspace{1em}
\end{abstract}

\section{Introduction}\label{introduction}

Artificial intelligence systems are no longer mere tools; they have become active participants in human cognitive processes. From diagnostic support in medicine to code generation in software engineering and literature synthesis in research, AI increasingly shapes how humans think, decide, and create. While this integration often yields impressive short-term performance gains, it also raises a fundamental question: does the AI truly \emph{amplify} human cognition, or does it gradually \emph{delegate} core reasoning tasks away from the human, leading to long-term dependence and skill erosion?

The distinction between cognitive amplification and cognitive delegation is not merely semantic. Amplification occurs when the human--AI team achieves results superior to what either could accomplish alone, while the human's independent capabilities remain intact or even improve over time. Delegation, by contrast, involves outsourcing significant portions of the cognitive workload to the AI, often resulting in hybrid outputs where the artificial component dominates and the human's unaided performance degrades with prolonged use.

Existing literature on automation bias, cognitive offloading, and extended cognition has documented these risks, yet few frameworks provide operational metrics to quantify and distinguish the two regimes in practice. Most evaluations still rely on aggregate accuracy or user satisfaction, metrics that fail to capture whether observed performance gains come at the cost of human cognitive sustainability. As a result, systems that appear effective in the short term may still conceal progressive dependence, loss of retained expertise, or only superficial forms of collaboration.

The motivating concern is therefore empirically grounded but operationally 
under-specified. Prior work gives reason to believe that AI assistance can change 
human judgment and reliance patterns; the present paper asks how such changes 
should be measured at the level of the human--AI system.

In this paper, we address this gap by proposing a compact evaluative framework built around four operational quantities. The Cognitive Amplification Index ($CAI^*$) measures collaborative gain relative to the best standalone agent. The Dependency Ratio ($D$) and Human Reliance Index ($HRI$) characterize the structural balance of contribution within hybrid outputs. Most critically, the Human Cognitive Drift Rate ($HCDR$) tracks changes in autonomous human performance over time through periodic evaluation when AI assistance is temporarily withheld. Taken together, these metrics define a low-dimensional evaluative space that makes visible the tension between short-term hybrid performance and long-term cognitive sustainability.

This perspective motivates a \emph{cognitive sustainability constraint}: hybrid performance $Q_{HA}$ should not be improved at the cost of degrading retained human capability over time. In formal terms, one should seek to maximize $Q_{HA}$ subject to $\text{HCDR} \geq 0$. When this condition is violated, a system may appear effective in the short term while remaining structurally dependent on the artificial component and progressively weakening the human's unaided capability.

To probe the discriminative capacity of the framework under controlled assumptions, we implement an agent-based simulation in NetLogo and evaluate three reliance regimes across multiple dependency--atrophy configurations. The results show that the proposed metrics distinguish degenerate delegation, interaction that is more capability-preserving than full delegation but weakly competitive, and intermediate boundary regimes that approach the AI baseline while remaining structurally dependent. Across the tested configurations, no regime achieves genuine amplification, and a constrained optimization over the atrophy parameter shows that even eliminating atrophy is not, by itself, sufficient to recover positive collaborative gain within the present model.

Additional parameter sweeps further indicate that this limitation is not merely
parametric: within the baseline interaction structure, positive collaborative
gain remains unattainable, motivating a structural extension in which an explicit
human--AI interaction term is introduced.

The main contributions of this paper are fourfold. First, we introduce a compact evaluative framework for distinguishing cognitive amplification from cognitive delegation in human--AI systems. Second, we formalize this distinction through four operational metrics---$CAI^*$, $D$, $HRI$, and $HCDR$---that jointly capture collaborative gain, dependency structure, and temporal change in unaided human capability. Third, we provide an initial structural validation of the framework through an agent-based simulation and a constrained optimization experiment, showing that reducing capability erosion alone is insufficient to recover genuine amplification within the baseline model. Fourth, we incorporate a structural extension showing that positive amplification becomes attainable only when the interaction model includes an explicit human--AI synergy term, suggesting that amplification is a property of the coupling architecture rather than a generic consequence of AI assistance.

The remainder of the paper is organized as follows. Section 2 reviews related work on human--AI collaboration, overreliance, cognitive offloading, and cognitively sustainable system design. Section 3 introduces the system model and formal definitions of the proposed metrics. Section 4 presents a conceptual phase diagram and discusses the resulting collaboration regimes. Section 5 develops design implications and provides an illustrative task-level example. Section 6 presents the agent-based simulation and experimental setup. Section 7 reports the main baseline empirical results and the atrophy-optimization experiment. Section 8 presents additional parameter sweeps and a structural extension examining the conditions under which amplification can become attainable. The paper then discusses broader implications, limitations, future directions, and concludes.

\section{Related Work}

\subsection{Human--AI collaboration and complementarity}

Recent work on human--AI collaboration emphasizes that the effectiveness of hybrid systems depends not only on the raw capabilities of the AI, but also on the quality of interaction, task structure, and the extent to which human and artificial agents contribute complementary information or capabilities \cite{kamar2016directions, gombolay2023humanai}. Research on complementarity in human--AI decision-making argues that productive collaboration emerges when differences in access to information, representational strengths, or error patterns can be productively combined \cite{bansal2021does} rather than when one agent merely replicates the function of the other.

Related studies further show that superior hybrid performance should not be taken for granted: meta-analytic evidence indicates that human–AI combinations often fail to outperform the best standalone agent and that performance gains are highly contingent on task and interaction design \cite{when-combinations-useful}. This makes the design of effective interaction protocols a central challenge rather than a secondary implementation detail. These findings motivate evaluation frameworks that go beyond raw hybrid accuracy. If performance gains are highly contingent on how the interaction is organized, then assessment must also examine \emph{how} those gains are achieved and whether they preserve an active human cognitive role.

\subsection{Overreliance, automation bias, and cognitive offloading}

A substantial body of work has documented that humans frequently over-rely on automated decision aids. Research on automation bias describes the tendency of users to assign excessive weight to automated recommendations, reduce independent verification, and inherit system errors---particularly in high-stakes or cognitively demanding tasks \cite{skitka1999automation, automation-bias}. More recent literature on AI overreliance extends these concerns to contemporary generative systems, showing that AI assistance can subtly alter judgment strategies, reduce scrutiny, and create a misleading appearance of robust human oversight \cite{overreliance-ai, dellacqua2023navigating}.
Recent experimental work further shows that users can internalize and reproduce the biases present in AI recommendations even when later making unaided judgments \cite{humans-inherit-ai-bias}.

These issues connect closely to the literature on cognitive offloading. Studies in this area demonstrate that external aids can boost immediate task performance while simultaneously reshaping how users remember, reason, and allocate mental effort \cite{risko2016, cognitive-offloading}. In the context of AI-assisted decision-making, this raises a critical possibility: a system may improve short-term hybrid performance while weakening the user's unaided capacity to perform comparable tasks. This temporal distinction---between immediate gains and long-term cognitive effects---lies at the core of the present paper.

\subsection{Human-centred AI and cognitively sustainable interaction}

Human-centred AI has increasingly been framed not only as a question of usability or explainability, but as a broader design commitment to preserving meaningful human agency, oversight, and participation \cite{designing-intelligent-organization}. From this perspective, effective human--AI interaction requires more than accurate model outputs: it demands interfaces and workflows that calibrate trust, support critique, expose uncertainty, and maintain the human's ability to reason independently about the task \cite{designing-ai-expertise}.

Nevertheless, current evaluation practices in human--AI systems continue to prioritize immediate hybrid performance. Far less attention has been devoted to whether interaction designs preserve---or erode---human competence over time. The present work contributes to this gap by introducing a metric framework that explicitly distinguishes performance gains reflecting genuine \emph{cognitive amplification} from those that mask progressive \emph{cognitive delegation}.

\subsection{Evaluation metrics for human--AI systems}

Several prior works have proposed computational frameworks and metrics for assessing human--AI collaboration \cite{kittur2020humanai}. These include measures of team performance, trust calibration, complementarity, and workload distribution. However, most existing metrics focus on immediate task outcomes or user satisfaction rather than on the temporal evolution of human capability under prolonged AI assistance \cite{impact-ai-cognition}.

The framework introduced in this paper extends this line of work by defining metrics that jointly capture collaborative gain, dependency structure, and cognitive drift over time. By introducing an explicit temporal dimension through the Human Cognitive Drift Rate ($HCDR$), the proposed framework enables longitudinal evaluation of whether AI assistance preserves or degrades autonomous human performance. This addresses a gap in current evaluation practice, where temporal effects on retained human capability are rarely measured or reported.

\subsection{Conceptual motivation}

The distinction proposed here also draws on broader perspectives concerning the nature of human cognition and the risks of treating artificial computation as a functional substitute for human reasoning \cite{solms-friston-2019, searle1980}. While this paper does not seek to resolve debates in the philosophy of mind \cite{byrne-kim-2012}, it acknowledges that human and artificial systems may produce similar task-level outputs while relying on fundamentally different underlying processes \cite{simon1996, clark1998}. This asymmetry motivates treating the preservation of human cognitive competence as a first-class design concern rather than as a secondary consideration. The proposed framework operationalizes this perspective by making cognitive sustainability an explicit evaluation criterion alongside immediate hybrid performance.

\subsection{Explainability}
AI systems should expose uncertainty, alternatives, and supporting evidence rather than presenting a single authoritative output for acceptance. From a cognitive perspective, explanations support abductive inference and allow users to test and revise hypotheses rather than merely consuming outputs \cite{lombrozo2012explanation}.
This suggests that explanation mechanisms should be evaluated not only in terms of immediate user satisfaction, but also by their impact on human reasoning strategies and long-term cognitive drift.

\section{System Model}

Consider a hybrid human--AI system $S$ composed of a human agent $H$ and an artificial agent $A$. We represent this coupling symbolically as
\begin{equation}
S = H + A,
\end{equation}
where the ``$+$'' symbol is not intended as a literal additive decomposition of intelligence, but as a compact notation for a coupled system whose effective behavior emerges from the interaction between both components.

Human intelligence involves forms of understanding, judgment, and self-monitoring that cannot be fully captured by any single performance metric. Nevertheless, to evaluate whether an AI system amplifies or displaces human cognitive function, an operational measure of task-relevant capability is required. We therefore define the effective problem-solving capacity of the system as
\begin{equation}
Q(S) = Q(H,A),
\end{equation}
where $Q$ denotes a domain-specific, operational measure of cognitive performance (e.g., accuracy, solution quality, decision correctness, or other task-appropriate indicators). This is explicitly \emph{not} a measure of general intelligence.

The standalone performance of the human is denoted
\begin{equation}
Q_H = Q(H),
\end{equation}
while the standalone performance of the AI agent is
\begin{equation}
Q_A = Q(A).
\end{equation}
The hybrid performance is then
\begin{equation}
Q_{HA} = Q(S) = Q(H,A).
\end{equation}

The central question addressed in this paper is not whether AI can fully reproduce human intelligence, but whether AI systems enhance human cognitive agency or progressively replace its practical exercise.
In the present simulation, the standalone AI performance $Q_A$ is operationalized as a fixed reliability parameter rather than as a task-adaptive model output.

\section{Cognitive Amplification}

In the ideal case, the AI functions as a genuine cognitive amplifier. The hybrid system exhibits synergy: machine assistance extends human problem-solving capacity without suppressing the active exercise of human judgment and self-monitoring.

This synergistic effect can be expressed conceptually as
\begin{equation}
Q_{HA} = Q_H + Q_A + \alpha \, Q_H Q_A,
\label{eq:synergy}
\end{equation}
where $\alpha > 0$ represents the strength of the productive interaction between human and AI. When $\alpha > 0$, the coupled system produces a positive interaction term beyond the sum of the isolated contributions.

This expression should be interpreted as an idealized model of human--AI synergy rather than a literal empirical law. In practice, performance measures are typically bounded and strongly task-dependent, and hybrid performance rarely scales in a strictly multiplicative fashion. For this reason, the framework developed in the following sections relies on relative performance metrics that compare the hybrid output against the relevant standalone baselines of $H$ and $A$ individually. These relative metrics allow us to quantify amplification without assuming any particular functional form for the interaction.

\section{Metrics for Human--AI Collaboration}

The system model defines three core performance quantities: $Q_H$, $Q_A$, and $Q_{HA}$. From these we derive four operational metrics that, when interpreted jointly, characterize the regime of human--AI interaction. No single metric is sufficient; the most informative assessment requires reading them together.

\subsection{Cognitive Amplification Index}

The Cognitive Amplification Index ($CAI^*$) quantifies the relative performance gain of the hybrid system over the best standalone agent:
\begin{equation}
CAI^* = \frac{Q_{HA} - \max(Q_H, Q_A)}{\max(Q_H, Q_A)}.
\end{equation}

A positive value indicates that the coupled human--AI system exceeds both standalone baselines. A zero value indicates that the hybrid system matches the best standalone agent without producing additional collaborative gain. A negative value indicates that integration fails to outperform the best standalone component.

A positive $CAI^*$ is therefore a necessary condition for cognitive amplification, but it is not sufficient on its own. Short-term collaborative gain may coexist with high dependency or declining unaided human capability. For this reason, $CAI^*$ must be interpreted jointly with the dependency and drift metrics defined below.

\subsection{Dependency Ratio and Human Reliance Index}

The Dependency Ratio ($D$) measures the extent to which hybrid performance is anchored in the AI component:
\begin{equation}
D = \frac{Q_A}{Q_{HA}}.
\end{equation}

High values of $D$ indicate strong AI dominance and limited human contribution, consistent with patterns of overreliance and automation bias.

The complementary Human Reliance Index is defined as
\begin{equation}
HRI = 1 - D = \frac{Q_{HA} - Q_A}{Q_{HA}}.
\end{equation}

In the present simulation, $Q_A$ is operationalized as a fixed baseline parameter representing AI reliability rather than a task-dependent measured performance. As a result, the Dependency Ratio should be interpreted as a comparative dominance indicator rather than as a strict mixture coefficient. In particular, values exceeding 1 may arise when observed hybrid performance falls below the nominal AI baseline, indicating ineffective integration rather than literal over-contribution of the AI component. Accordingly, in the present simulation $HRI$ should be interpreted as a relative human-contribution indicator rather than as a bounded proportion; when $Q_{HA} < Q_A$, negative $HRI$ values do not imply ``negative reliance'' in a literal sense, but rather that the hybrid system underperforms the nominal AI baseline.

\begin{table}[htbp]
\centering
\begin{tabular}{ccc}
\toprule
Range of $D$ & Range of $HRI$ & Qualitative regime \\
\midrule
$< 0.5$      & $> 0.5$        & Human-dominant \\
$0.5$--$0.8$ & $0.2$--$0.5$   & Balanced collaboration \\
$0.8$--$1.0$ & $0$--$0.2$     & AI-dominated (risk of delegation) \\
$> 1.0$      & $< 0$          & Hybrid underperforms the nominal AI baseline \\
\bottomrule
\end{tabular}
\caption{Heuristic interpretive regimes based on the Dependency Ratio ($D$) and Human Reliance Index ($HRI$). In the present simulation, $Q_A$ is treated as a fixed baseline, so $D>1$ and $HRI<0$ may arise when the hybrid system performs below the nominal AI baseline.}
\label{tab:dependency}
\end{table}

These threshold ranges are heuristic and primarily intended for interpretive guidance. In the present simulation, empirical values of $D$ and $HRI$ should therefore be read comparatively rather than as hard-bounded contribution shares.

\subsection{Human Cognitive Drift Rate}

While $CAI^*$, $D$, and $HRI$ describe the system state at a given moment, the Human Cognitive Drift Rate ($HCDR$) captures the temporal evolution of unaided human performance:
\begin{equation}
HCDR = \frac{Q_H(t_2) - Q_H(t_1)}{t_2 - t_1},
\end{equation}
where $Q_H(t)$ is the human's performance measured without AI assistance through periodic ``AI-off'' assessments.

\begin{itemize}
\item $HCDR \geq 0$ (amplification regime): unaided human performance is preserved or improves over time.
\item $HCDR < 0$ (delegation regime): unaided human performance deteriorates, consistent with cognitive offloading and deskilling effects.
\end{itemize}

$HCDR$ is the metric most directly linked to long-term cognitive sustainability.

\subsection{Regimes of Human--AI Collaboration}

The metrics above define a compact state space for human--AI systems. In particular, the pair $(D, CAI^*)$ provides a useful low-dimensional representation, where $CAI^*$ captures performance synergy and $D$ reflects the balance of contribution.

Figure~\ref{fig:phase_diagram} presents a conceptual phase diagram in this space. The horizontal axis represents the Dependency Ratio $D$ (increasing AI dominance), while the vertical axis shows the Cognitive Amplification Index $CAI^*$. The Human Cognitive Drift Rate ($HCDR$) acts as a third dimension that determines whether a given regime is cognitively sustainable in the long term.

\begin{figure}[htbp]
\centering
\resizebox{0.92\textwidth}{!}{%
\begin{tikzpicture}[
    >=Latex,
    font=\small
]

\draw[->, thick] (0,0) -- (10.5,0)
    node[right, align=center] {Dependency Ratio $D$\\(AI dominance)};
\draw[->, thick] (0,0) -- (0,7.2)
    node[above, align=center] {Cognitive Amplification Index\\$CAI^*$};

\node[below left] at (0,0) {$0$};
\node[left] at (0,3.5) {$0$};

\draw[dashed] (5,0) -- (5,7);
\draw[dashed] (0,3.5) -- (10,3.5);

\node[
    align=center,
    text width=3.4cm,
    fill=white,
    draw,
    rounded corners,
    inner sep=6pt
] at (2.5,5.4)
{ \textbf{Cognitive Amplification}\\
Balanced collaboration\\
$CAI^* > 0$, moderate $D$\\
$HCDR \geq 0$ };

\node[
    align=center,
    text width=3.5cm,
    fill=white,
    draw,
    rounded corners,
    text=red!80!black,
    inner sep=6pt
] at (7.5,5.4)
{ \textbf{Automation Trap}\\
High short-term performance\\
High $D$, positive $CAI^*$\\
Negative $HCDR$ };

\node[
    align=center,
    text width=3.4cm,
    fill=white,
    draw,
    rounded corners,
    inner sep=6pt
] at (2.5,1.5)
{ \textbf{Human-Dominant Regime}\\
Low dependency\\
Limited AI contribution\\
Little or no amplification };

\node[
    align=center,
    text width=3.6cm,
    fill=white,
    draw,
    rounded corners,
    inner sep=6pt
] at (7.5,1.5)
{ \textbf{Ineffective Automation}\\
High dependency\\
$CAI^* < 0$ or fragile gains\\
Poor hybrid integration };

\draw[->, thick, red!70!black] (9.2,6.2) -- (9.2,4.7);
\node[right, text=red!80!black, align=left] at (9.35,5.45) {$HCDR < 0$};

\end{tikzpicture}%
}
\caption[Conceptual phase diagram of human--AI collaboration regimes]{Conceptual phase diagram of human--AI collaboration regimes. The horizontal axis represents the Dependency Ratio $D$, and the vertical axis represents the Cognitive Amplification Index $CAI^*$. The upper-right region corresponds to the \emph{automation trap}, where strong short-term hybrid performance coexists with high AI dominance and negative human cognitive drift.}
\label{fig:phase_diagram}
\end{figure}

These regimes demonstrate that maximizing short-term hybrid performance $Q_{HA}$ does not automatically guarantee cognitively sustainable collaboration. A system may exhibit positive $CAI^*$ and appear efficient while still operating in the automation trap if $HCDR < 0$. Conversely, a more balanced regime with moderate $CAI^*$ but non-negative $HCDR$ may be preferable when long-term human expertise is a priority.

\section{Design Implications}

The metric framework introduced above has immediate implications for the design and evaluation of human--AI systems. In safety-critical and knowledge-intensive domains, the objective should extend beyond maximizing short-term hybrid performance. The more demanding target is to support collaboration that preserves human capability over time, limits structural dependence on the artificial component, and, ideally, produces genuine collaborative gain.

From this perspective, three design goals become especially important:
\begin{itemize}
\item preserving or improving autonomous human performance over time ($HCDR \ge 0$),
\item avoiding regimes of excessive AI dominance, reflected in persistently high $D$ and low HRI,
\item and creating conditions under which the hybrid system can exceed the best standalone baseline ($CAI^* > 0$).
\end{itemize}

The present results do not, by themselves, validate specific interface prescriptions. However, the framework suggests several plausible design directions for moving systems away from delegation and toward amplification.

First, interaction protocols should keep the human actively inside the reasoning loop rather than positioning the AI as a final-answer oracle. Mechanisms such as mandatory human-first attempts, explicit hypothesis generation, or explanation-before-revelation may help reduce passive acceptance and preserve retained capability over time.\cite{cognitive-offloading,designing-ai-expertise,ai-pitfalls}

Second, AI systems should expose uncertainty, alternatives, and supporting evidence rather than presenting a single authoritative output for acceptance. This may help prevent structural overreliance and maintain a visible human contribution to the hybrid result.\cite{human-ai-uq,uncertainty-explainability,ai-cognitive-implications}

Third, cognitively sustainable deployment likely requires explicit capability-preservation mechanisms, including periodic AI-off evaluation, autonomous practice blocks, or training-oriented assistance modes. The optimization experiment reported in this paper suggests that preserving capability matters, but also that capability preservation alone is not sufficient to recover genuine amplification under the present interaction dynamics.

From an engineering standpoint, these implications suggest that human--AI systems should incorporate telemetry not only for output quality, but also for collaborative gain, dependency structure, and human cognitive drift. Such instrumentation would make it possible to detect when a system is moving from acceptable assistance toward hidden delegation, and would provide a basis for iterative redesign of both workflows and interaction policies.\cite{designing-intelligent-organization,automation-bias-ai-act}
In regulatory contexts such as the EU AI Act, these concerns are amplified by the legal requirement for meaningful human oversight, which can be undermined by automation bias \cite{automation-bias-ai-act}.

\section{\texorpdfstring{Illustrative Measurement of $Q$}{Illustrative Measurement of Q}}

The framework introduced in this paper does not require a universal measure of intelligence. Instead, $Q$ denotes the effective problem-solving capability of a system within a specific task domain. In practice, $Q$ may be approximated using task-level performance measures such as decision accuracy, solution quality, task completion time, or error rates in controlled evaluation settings.

To illustrate the framework, consider a simplified engineering diagnostic task in which an operator must identify the root cause of a system anomaly from sensor signals. Suppose the following performance levels are observed:

\begin{center}
\begin{tabular}{lc}
\toprule
System & Diagnostic Accuracy \\
\midrule
Human alone ($Q_H$) & 0.70 \\
AI alone ($Q_A$) & 0.80 \\
Human + AI ($Q_{HA}$) & 0.92 \\
\bottomrule
\end{tabular}
\end{center}

The Cognitive Amplification Index is then

\begin{equation}
CAI^* = \frac{Q_{HA} - \max(Q_H, Q_A)}{\max(Q_H, Q_A)}
      = \frac{0.92 - 0.80}{0.80}
      \approx 0.15.
\end{equation}

The Dependency Ratio becomes

\begin{equation}
D = \frac{Q_A}{Q_{HA}} = \frac{0.80}{0.92} \approx 0.87,
\end{equation}

and the complementary Human Reliance Index is therefore

\begin{equation}
HRI = 1 - D \approx 0.13.
\end{equation}

In this example, the hybrid system clearly outperforms the best standalone component ($CAI^* > 0$), but the relatively high value of $D$ indicates that performance remains strongly anchored to the AI baseline. This illustrates an important point of the framework: positive collaborative gain does not by itself guarantee a balanced or cognitively sustainable regime. The same logic applies across domains such as medical decision support, industrial diagnosis, software engineering, or safety monitoring, provided that $Q$ is instantiated through an appropriate task-specific performance measure.

\section{Agent-Based Simulation for Structural Validation}

To explore the dynamic consequences of cognitive amplification and cognitive delegation without exposing human participants to potentially harmful deskilling conditions, we implement an agent-based simulation in NetLogo. 
Agent-based modeling is well suited to this problem because it allows macro-level patterns of dependence, collaborative gain, cognitive drift, and adaptation failure to emerge from simple micro-level behavioral rules defined at the level of individual agents. NetLogo further provides a practical environment for large-scale stochastic experimentation, including systematic execution through BehaviorSpace.
The mechanisms implemented in the simulation are not intended to prove that 
delegation occurs in real users; rather, they instantiate, in simplified 
computational form, mechanisms already motivated by prior work on automation bias, 
cognitive offloading, and overreliance.

\subsection{Purpose of the simulation}

The purpose of the simulation is not to reproduce human cognition in a psychologically complete sense, but to test whether the metric framework introduced in this paper can recover qualitatively distinct regimes of human--AI interaction under controlled assumptions. In particular, the simulation is designed to examine whether strong short-term hybrid performance can coexist with negative human cognitive drift, thereby producing the automation trap identified in the conceptual phase diagram.

The simulation therefore operationalizes the distinction between \emph{cognitive amplification} and \emph{cognitive delegation} in a stylized environment where agents face a stream of randomly generated tasks, decide whether to rely on an AI oracle, and update their internal skill state over time. The main observable of interest is whether repeated low-effort delegation improves immediate productivity while degrading the agent's autonomous capability when AI support is intermittently removed.

A further objective is to distinguish between competence that is merely preserved within a familiar task distribution and competence that remains robust under novelty. In many practical settings, the key risk of excessive delegation is not an immediate collapse in routine performance, but a progressive loss of adaptability when task structure changes. For this reason, the simulation also evaluates whether agents retain the ability to cope with previously unseen task configurations or shifts in the distribution of problem types after extended AI reliance.

The simulation should therefore be read as a stylized computational testbed rather 
than as a realistic psychological model of human cognition. Its role is not to 
estimate real-world rates of cognitive atrophy, overreliance, or skill retention, 
nor to claim that human learning and deskilling follow the particular update rules 
implemented in the model. Instead, the simulation serves a narrower methodological 
purpose: to test whether the proposed metrics behave discriminatively under 
controlled and transparent assumptions about reliance, shallow learning under AI 
assistance, disuse-related capability erosion, and task-distribution shift.

In this sense, the relevant form of validity is structural and diagnostic rather 
than predictive. The model introduces minimal mechanisms that are consistent with 
known concerns in automation bias, cognitive offloading, and human--AI 
overreliance, and then asks whether the metric framework can separate interaction 
regimes that would be difficult to distinguish using hybrid performance alone. 
The simulation therefore evaluates the construct behavior of the framework: 
whether $CAI^*$, $D$, $HRI$, and $HCDR$ jointly distinguish amplification-like, 
delegation-like, and boundary regimes under controlled conditions.

Accordingly, the results should not be interpreted as empirical evidence that a 
specific human population would exhibit the exact trajectories reported here. They 
should instead be interpreted as an operational demonstration that the framework 
can expose differences between short-term hybrid output, structural dependence, 
retained autonomous capability, and cognitive drift. Human-subject studies remain 
necessary to calibrate the numerical values of these quantities in real settings; 
however, such studies are logically subsequent to the present contribution, whose 
aim is to define the evaluative quantities and demonstrate their discriminative 
use in a reproducible computational environment.

\subsection{Agents, tasks, and skills}

Each agent represents a generic problem-solver rather than a novice learner. Agents are initialized with a non-zero baseline competence, reflecting the assumption that real operators typically begin with an already acquired set of task-relevant abilities. Formally, each agent $i$ is associated with a skill vector
\[
\mathbf{S}_i(t) = \left[s_{i1}(t), s_{i2}(t), \dots, s_{ik}(t)\right], \qquad s_{ij}(t) \in [0,1],
\]
where each component represents proficiency in one abstract cognitive skill dimension.

The initial condition is chosen such that agents already master part of the task space at the start of the simulation. This prevents trivial dependence on AI for tasks that are already familiar and forces the relevant dynamic to arise from exploration into less familiar regions of the problem space rather than from simple repetition of known routines.

Each task is represented as a requirement vector
\[
\mathbf{R}(t) = \left[r_1(t), r_2(t), \dots, r_k(t)\right], \qquad r_j(t) \in [0,1],
\]
together with a scalar difficulty parameter
\[
C(t) \in [0,1].
\]
Tasks are sampled randomly at each time step from a distribution over problem types and difficulty levels. In practice, the simulator uses a small set of abstract task families---analytical, diagnostic, sequential, and mixed---where each family activates a different subset of skill dimensions. This family-based representation preserves codable structure while allowing controlled shifts in the task distribution, thereby distinguishing performance on familiar tasks from performance under novelty or partial domain transfer.

\subsection{Task--skill mismatch, effort, and perturbation}

For each agent and task, the simulation computes a task--skill mismatch that approximates how demanding the task is relative to the agent's current internal capabilities:
\[
M_i(t) = \frac{1}{k} \sum_{j=1}^{k} \max\{0,\, r_j(t) - s_{ij}(t)\}.
\]
This quantity is low when the task lies within the agent's already mastered region and increases when the task requires capabilities that the agent has not yet sufficiently developed.

In addition to nominal task difficulty, the model includes a stochastic perturbation term representing unforeseen complications, contextual irregularities, or minor deviations from routine execution:
\[
\widetilde{C}(t) = C(t) + \epsilon(t),
\]
where $C(t) \in [0,1]$ is nominal difficulty and $\epsilon(t)$ is a bounded random perturbation. The perceived cognitive effort required to engage the task autonomously is then modeled as
\[
E_i(t) = \lambda_M M_i(t) + \lambda_C \widetilde{C}(t),
\]
with $\lambda_M, \lambda_C > 0$ as tunable parameters.

This construction is important for interpretation. The central issue is not whether agents can repeatedly solve already mastered tasks, but whether they continue to invest effort in cognitively demanding tasks that would expand or preserve competence. The effort term therefore acts as the main pressure driving agents toward either exploration or delegation, while the perturbation term ensures that routine competence is never evaluated under unrealistically frictionless conditions.

\subsection{Learning dynamics and cognitive atrophy}

The simulation is intentionally designed so that agents do not explicitly ``solve'' tasks in a symbolic sense. Instead, tasks act as opportunities for skill acquisition or skill erosion depending on the mode of interaction.

When the agent engages the task without AI support, the relevant skill dimensions are updated strongly. Let $\mathcal{A}(t)$ denote the subset of activated skill dimensions for the current task. Then the autonomous learning update is
\[
s_{ij}(t+1) = s_{ij}(t) + \alpha_{\mathrm{self}} \, (1 - s_{ij}(t)) \, r_j(t),
\qquad j \in \mathcal{A}(t),
\]
where $\alpha_{\mathrm{self}}$ is the self-learning rate. The factor $(1 - s_{ij}(t))$ induces diminishing returns near saturation and prevents unrealistic linear growth.

When the agent uses AI assistance, the same task produces only weak learning:
\[
s_{ij}(t+1) = s_{ij}(t) + \alpha_{\mathrm{AI}} \, (1 - s_{ij}(t)) \, r_j(t),
\qquad j \in \mathcal{A}(t),
\]
with
\[
0 < \alpha_{\mathrm{AI}} \ll \alpha_{\mathrm{self}}.
\]
Thus, AI use is not modeled as yielding zero learning, but as producing shallow acquisition relative to autonomous engagement.

To capture disuse effects, repeated AI reliance may also induce a downward drift in capability through an atrophy term with rate $\delta > 0$. In the present interpretation, atrophy is not merely a reduction in an abstract internal variable. Its practical significance is that it reduces autonomous robustness when task demands increase, when familiar tasks present unexpected complications, or when novel task structures are encountered. In this sense, cognitive atrophy represents a loss of resilience: agents may continue to function acceptably under routine conditions while becoming progressively less able to cope with perturbation, increased difficulty, or domain shift.

The central hypothesis is therefore that prolonged AI reliance may preserve acceptable local performance while weakening the deeper skill structure required for autonomous adaptation. In that case, the effect of delegation may remain partially hidden as long as the task environment stays familiar, and may become visible only when the agent is evaluated under altered or perturbed task conditions.

\subsection{Productivity, hybrid performance, and robustness}

Immediate task performance is modeled separately from long-term skill adaptation. If the agent uses the AI oracle, the task is completed with high short-term effectiveness, yielding strong instantaneous productivity. If the agent engages autonomously, immediate performance depends on its current skill profile, the mismatch of the task, and the effective difficulty of the current realization.

This separation allows the simulation to reproduce the central tension of the paper: a regime may exhibit strong short-term output while silently accumulating negative drift in autonomous capability.

At the population level, the simulation records mean autonomous skill level, AI usage frequency, immediate task productivity, estimated hybrid performance $Q_{HA}(t)$, autonomous performance $Q_H(t)$ measured during AI-off blocks, and autonomous performance under perturbed and novel task distributions. These observables allow direct estimation of the framework metrics introduced earlier, including $CAI^*$, $D$, and especially $HCDR$.

\subsection{AI-off evaluation and robustness probes}

To estimate Human Cognitive Drift Rate, the simulation periodically enters evaluation windows in which AI assistance is temporarily disabled. During these windows, autonomous performance is measured directly using the current skill vectors under the same task distribution used during normal interaction. Let $Q_H(t_m)$ denote the average autonomous performance measured at evaluation time $t_m$. The drift rate is approximated by
\[
HCDR \approx \frac{Q_H(t_{m+1}) - Q_H(t_m)}{\Delta t},
\]
where $\Delta t$ corresponds to the interval between evaluation windows.

A negative value indicates that repeated reliance on AI has degraded unaided capability over time, even if hybrid productivity remains high during normal operation. This AI-off protocol is critical because it distinguishes genuine cognitive amplification from operational dependence masked by strong AI performance.

In addition, the simulation includes two forms of robustness probe. \emph{Perturbation probes} introduce stochastic increases in difficulty for otherwise familiar task structures, producing a measure $Q_H^{\mathrm{pert}}(t)$. \emph{Novelty probes} sample tasks from shifted or reweighted task-family distributions, producing a measure $Q_H^{\mathrm{novel}}(t)$. These probes are executed periodically throughout the simulation rather than being restricted to a final evaluation stage. This allows the analysis to distinguish routine competence, robustness to variability within the same domain, and adaptive competence under distributional shift.

\subsection{Experimental phases}

The NetLogo simulation is organized into three phases.

\subsubsection{Phase 1: Baseline without AI.}
Agents interact with randomly generated tasks using only autonomous learning. This establishes an initial trajectory of skill acquisition and provides a reference level for $Q_H$.

\subsubsection{Phase 2: AI-assisted interaction.}
AI assistance becomes available, and agents interact with the AI oracle according to their assigned reliance regime. Periodic AI-off evaluation windows are introduced during this phase, enabling direct estimation of Human Cognitive Drift Rate ($HCDR$).

\subsubsection{Phase 3: Shifted-environment stress test.}
In the final phase, the task distribution is modified to increase the prevalence of more complex or composite task families, effectively introducing a domain shift. AI assistance remains available, but the environment becomes less routine and more demanding in terms of generalization and transfer. Autonomous robustness is assessed through continued AI-off evaluation together with perturbation and novelty probes.

This three-phase structure makes it possible to distinguish baseline learning dynamics without AI, short-term hybrid performance under AI assistance, and longer-term robustness under distributional shift.

\subsection{Experimental protocol}

The experiments were implemented in NetLogo and executed through BehaviorSpace. We evaluated three reliance regimes as separate experimental modes: full delegation ($p_{\mathrm{AI}}=1.0$), minimal AI ($p_{\mathrm{AI}}=0.0$), and mixed reliance ($p_{\mathrm{AI}}=0.5$). These modes allow a clean comparison between maximal delegation, near-autonomous operation, and intermediate reliance under otherwise identical environmental conditions.

To probe different balances between capability erosion and dependency reinforcement, we studied three parameter configurations:
\begin{itemize}
\item \textbf{P0:} dependency-use-sensitivity $=0.2$, atrophy-delta $=0.004$
\item \textbf{P1:} dependency-use-sensitivity $=0.6$, atrophy-delta $=0.003$
\item \textbf{P2:} dependency-use-sensitivity $=0.4$, atrophy-delta $=0.002$
\end{itemize}

Unless otherwise stated, all remaining parameters were held fixed across experiments, including $N=1000$, $k=6$, $\alpha_{\mathrm{self}}=0.05$, and $\alpha_{\mathrm{AI}}=0.00105$. Each condition was repeated over 20 random seeds, and the reported metrics were aggregated over a final evaluation window rather than extracted from a single terminal step. For each run, we recorded the four principal quantities introduced in this paper---$CAI^*$, $D$, HRI, and HCDR---together with supporting observables such as autonomous performance, hybrid performance, retained skill, AI usage rate, and robustness under perturbation and novelty probes.

\section{Results}

All results in this section should be interpreted as outcomes of the stylized 
computational model described above. They are not intended as estimates of human 
deskilling rates, but as evidence that the proposed metric framework can 
discriminate among qualitatively different interaction regimes under controlled 
assumptions.

We evaluated three experimental configurations, denoted P0, P1, and P2, each representing a different combination of dependency-use sensitivity and atrophy rate within the agent-based model. Within each configuration, we compared three reliance regimes: full delegation, minimal AI, and mixed reliance. All reported values are averaged over 20 random seeds.

Across all three configurations, the three reliance regimes separate clearly. Full delegation consistently collapses to a degenerate AI-dominated regime: AI usage saturates, retained human skill becomes minimal, and hybrid performance becomes indistinguishable from the AI baseline. Minimal AI preserves the highest autonomous human capability and the highest retained skill, but its hybrid performance remains substantially below the AI baseline, yielding strongly negative collaborative gain. Mixed reliance occupies an intermediate position in all configurations: it preserves substantially more human capability than full delegation, but still fails to produce genuine amplification, with $CAI^*<0$ and $D>1$ across seeds.

Under P2 (dependency-use-sensitivity $=0.4$, atrophy-delta $=0.002$), the mixed regime behaves as a boundary case. It retains moderate skill and autonomous performance while achieving hybrid performance close to the AI baseline. However, this apparent stability does not translate into positive collaborative gain: $CAI^*$ remains slightly negative and the dependency ratio remains above 1.

Under P1 (dependency-use-sensitivity $=0.6$, atrophy-delta $=0.003$), stronger dependency feedback pushes the mixed regime even closer to the AI baseline. In this setting, mixed reliance remains distinct from full delegation, but its hybrid behavior becomes more tightly coupled to AI performance, further suppressing the possibility of positive collaborative gain.

Under P0 (dependency-use-sensitivity $=0.2$, atrophy-delta $=0.004$), weaker dependency feedback reduces the aggressiveness of delegation, but higher atrophy still prevents amplification. In this configuration, mixed reliance uses AI less aggressively than in P1 or P2 and remains farther from the AI baseline, yet it still fails to achieve $CAI^*>0$.

Taken together, the experiments show that improved hybrid performance does not by itself imply cognitive amplification. Across all tested configurations, none of the reliance regimes achieves positive collaborative gain.

\begin{table}[htbp]
\centering
\small
\setlength{\tabcolsep}{4pt}
\resizebox{\textwidth}{!}{%
\begin{tabular}{
@{}
l
S[table-format=-1.4]
S[table-format=1.4]
S[table-format=1.4]
S[table-format=1.4]
S[table-format=-1.4]
S[table-format=1.4]
S[table-format=-1.4]
S[table-format=1.4]
@{}
}
\toprule
& \multicolumn{2}{c}{$CAI^*$}
& \multicolumn{2}{c}{$D$}
& \multicolumn{2}{c}{$HRI$}
& \multicolumn{2}{c}{$HCDR$} \\
\cmidrule(lr){2-3}
\cmidrule(lr){4-5}
\cmidrule(lr){6-7}
\cmidrule(lr){8-9}
Configuration (mixed regime) & {Mean} & {Std} & {Mean} & {Std} & {Mean} & {Std} & {Mean} & {Std} \\
\midrule
P0 $(0.2,\;0.004)$ & -0.0829 & 0.0039 & 1.0920 & 0.0050 & -0.0920 & 0.0050 & 0.0001 & 0.0037 \\
P1 $(0.6,\;0.003)$ & -0.0055 & 0.0012 & 1.0056 & 0.0012 & -0.0056 & 0.0012 & 0.0000 & 0.0037 \\
P2 $(0.4,\;0.002)$ & -0.0276 & 0.0031 & 1.0290 & 0.0030 & -0.0290 & 0.0030 & 0.0001 & 0.0037 \\
\bottomrule
\end{tabular}
}
\caption{Main framework metrics for the mixed-reliance regime across the three experimental configurations. Values are mean and standard deviation over 20 random seeds.}
\label{tab:mixed-configs}
\end{table}

\begin{table}[htbp]
\centering
\small
\setlength{\tabcolsep}{4pt}
\begin{tabular}{
@{}
l
S[table-format=-1.4]
S[table-format=1.4]
S[table-format=-1.4]
S[table-format=1.4]
S[table-format=-1.4]
S[table-format=1.4]
S[table-format=-1.4]
S[table-format=1.4]
@{}
}
\toprule
& \multicolumn{2}{c}{$CAI^*$}
& \multicolumn{2}{c}{$D$}
& \multicolumn{2}{c}{$HRI$}
& \multicolumn{2}{c}{$HCDR$} \\
\cmidrule(lr){2-3}
\cmidrule(lr){4-5}
\cmidrule(lr){6-7}
\cmidrule(lr){8-9}
Regime (P2) & {Mean} & {Std} & {Mean} & {Std} & {Mean} & {Std} & {Mean} & {Std} \\
\midrule
Full delegation & 0.0000 & 0.0000 & 1.0000 & 0.0000 & 0.0000 & 0.0000 & -0.0004 & 0.0039 \\
Minimal AI      & -0.2650 & 0.0190 & 1.4810 & 0.0420 & -0.4810 & 0.0420 & 0.0001 & 0.0038 \\
Mixed reliance  & -0.0276 & 0.0031 & 1.0290 & 0.0030 & -0.0290 & 0.0030 & 0.0001 & 0.0037 \\
\bottomrule
\end{tabular}
\caption{Framework metrics across reliance regimes under configuration P2. Values are mean and standard deviation over 20 random seeds.}
\label{tab:p2-regimes}
\end{table}

\subsection{Interpretation}

Taken together, the simulation results demonstrate that the proposed metric framework can discriminate among qualitatively distinct interaction regimes within the modeled environment. The NetLogo simulation shows that the four quantities introduced in this paper are sufficient to distinguish several practically relevant human--AI regimes, including degenerate AI-dominated delegation, human-preserving but weakly competitive interaction, and intermediate boundary regimes that approach the AI baseline while remaining structurally dependent.

Most importantly, the experiments show that strong hybrid performance is not enough to establish genuine cognitive amplification. A regime may exhibit less capability degradation than full delegation and still fail to achieve positive collaborative gain. In our simulations, the mixed-reliance regime repeatedly appears as such a boundary case: it can remain substantially more human-preserving than full delegation while still exhibiting $CAI^*<0$ and $D>1$.

This is precisely the distinction the framework is designed to detect. Within the present model, the results support the central evaluative claim: hybrid performance alone does not guarantee cognitive sustainability. This suggests that evaluation of human--AI systems should not stop at observed hybrid performance, but must also account for dependency structure, retained human capability, and cognitive drift over time~\cite{designing-intelligent-organization,cognitive-offloading,ai-cognitive-implications}.

\subsection{Optimization experiment: reducing effective atrophy}

To test whether the proposed framework can support concrete design improvement, we performed a targeted optimization experiment over a single parameter: the atrophy rate $\delta$ (implemented as \texttt{atrophy-delta}). This parameter was selected because it has a direct intervention interpretation. In practical terms, lowering effective atrophy may correspond to measures such as mandatory autonomous practice, periodic AI-off assessment, scaffolded assistance, or training protocols designed to convert AI use into retained human capability rather than passive substitution.

The experiment was implemented directly in NetLogo BehaviorSpace. All simulation parameters were held fixed except $\delta$, which was varied over a predefined search range under the mixed-reliance regime. The purpose of the experiment was to determine whether, within the present interaction model, controlling capability erosion alone is sufficient to recover genuine cognitive amplification under the mixed-reliance regime. The constrained search procedure is described in Appendix~\ref{app:optimization}.

The search was conducted under configuration P2, which is the most favorable of the three baseline configurations for this one-parameter study, as it combines reduced atrophy with only moderate dependency reinforcement. If reducing atrophy alone were sufficient to recover genuine amplification, P2 is the most plausible tested setting in which such an effect would be expected to appear.

The evaluation criterion was defined in terms of the framework itself. In particular, we asked whether there exists a value of $\delta$ such that the resulting system satisfies positive collaborative gain ($CAI^* > 0$) while preserving human capability over time ($HCDR \ge 0$). This turns the optimization into a constrained search over the space of capability-preserving interventions.

Table~\ref{tab:atrophy-optimization} summarizes the outcome of the optimization experiment. The comparison is between the baseline mixed-reliance configuration and the best value of $\delta$ found by the search. The final column reports whether amplification, in the strict sense adopted in this paper, is attainable by acting on atrophy alone.

\begin{table}[ht]
\centering
\small
\setlength{\tabcolsep}{4pt}
\begin{tabular}{
@{}
l
S[table-format=1.4]
S[table-format=-1.4]
S[table-format=1.4]
S[table-format=1.4]
l
@{}
}
\toprule
Condition & {$\delta$} & {$CAI^*$} & {$D$} & {$HCDR$} & {Amplification} \\
\midrule
Baseline mixed & 0.0020 & -0.0276 & 1.0291 & 0.0001 & No \\
Best found     & 0.0000 & -0.0207 & 1.0218 & 0.0001 & No \\
\bottomrule
\end{tabular}
\caption{Best feasible result from the constrained search over $\delta$ under mixed reliance. Even at zero atrophy, reducing capability erosion alone does not produce genuine amplification.}
\label{tab:atrophy-optimization}
\end{table}

The full grid-search results for the atrophy-parameter optimization are reported in Appendix~\ref{app:atrophy-full-results}. These results confirm a monotonic improvement as $\delta$ is reduced, but also show that even the limiting case $\delta=0$ does not produce genuine amplification.

\begin{figure}[ht]
\centering
\begin{tikzpicture}
\begin{axis}[
    width=0.47\textwidth,
    height=0.30\textwidth,
    xlabel={Atrophy rate $\delta$},
    ylabel={$CAI^*$},
    xmin=0, xmax=0.0041,
    x dir=reverse,
    ymin=-0.040, ymax=0.000,
    xtick={0.004,0.003,0.002,0.001,0.0005,0},
    xticklabel style={font=\scriptsize},
    yticklabel style={font=\scriptsize},
    xticklabels={0.004,0.003,0.002,0.001,5e-4,0},
    ytick={-0.04,-0.03,-0.02,-0.01,0},
    scaled x ticks=false,
    scaled y ticks=false,
    grid=major,
    mark size=2pt,
    title={Collaborative gain}
]
\addplot+[mark=*] coordinates {
    (0.0040,-0.0351)
    (0.0035,-0.0343)
    (0.0030,-0.0333)
    (0.0025,-0.0309)
    (0.0020,-0.0276)
    (0.0015,-0.0237)
    (0.0010,-0.0219)
    (0.0005,-0.0209)
    (0.0000,-0.0207)
};
\addplot[dashed] coordinates {(0,0) (0.0041,0)};
\end{axis}
\end{tikzpicture}
\hfill
\begin{tikzpicture}
\begin{axis}[
    width=0.47\textwidth,
    height=0.30\textwidth,
    xlabel={Atrophy rate $\delta$},
    ylabel={$D$},
    xmin=0, xmax=0.0041,
    x dir=reverse,
    ymin=1.020, ymax=1.040,
    xtick={0.004,0.003,0.002,0.001,0.0005,0},
    xticklabel style={font=\scriptsize},
    yticklabel style={font=\scriptsize},
    xticklabels={0.004,0.003,0.002,0.001,5e-4,0},
    ytick={1.02,1.025,1.03,1.035,1.04},
    scaled x ticks=false,
    scaled y ticks=false,
    grid=major,
    mark size=2pt,
    title={Dependency ratio}
]
\addplot+[mark=square*] coordinates {
    (0.0040,1.0374)
    (0.0035,1.0364)
    (0.0030,1.0354)
    (0.0025,1.0327)
    (0.0020,1.0291)
    (0.0015,1.0250)
    (0.0010,1.0231)
    (0.0005,1.0221)
    (0.0000,1.0218)
};
\addplot[dashed] coordinates {(0,1.0) (0.0041,1.0)};
\end{axis}
\end{tikzpicture}
\caption{Effect of reducing the atrophy parameter \texorpdfstring{$\delta$}{delta} under mixed reliance. Moving from left to right corresponds to lower atrophy. Lower atrophy improves collaborative gain and reduces dependency, but even zero atrophy does not produce genuine amplification: \texorpdfstring{$CAI^*$}{CAI*} remains below zero and \texorpdfstring{$D$}{D} remains above 1 throughout the explored range.}
\label{fig:atrophy-optimization}
\end{figure}

This experiment isolates one practically meaningful design lever. Its purpose is not to solve the full human--AI design problem, but to determine whether capability-preserving intervention alone can be sufficient. The negative result indicates that genuine amplification requires coordinated control of multiple mechanisms rather than isolated mitigation of cognitive erosion.

\section{Additional Parameter Sweeps and Structural Extension}
\label{sec:structural-extension}

The baseline experiments and atrophy-optimization study show that, within the original interaction model, reducing capability erosion is not sufficient to produce positive collaborative gain. To examine whether this failure was merely due to the selected parameter values, we performed additional parameter sweeps over the main reliance, dependency, atrophy, and AI-performance parameters. The purpose of these sweeps was to determine whether some region of the original parameter space could yield $CAI^*>0$ without changing the model structure.

Across these additional sweeps, positive cognitive amplification did not emerge. This negative result suggested that the limitation was not only parametric, but structural: in the baseline assisted-output rule, AI use effectively replaces autonomous human contribution with an AI-mediated output anchored to the AI baseline. Under such a rule, the model can represent preservation, erosion, or dependency, but it has no explicit mechanism by which retained human competence can generate additional collaborative gain beyond the best standalone component.

\subsection{Structural ceiling in the baseline model}

In the baseline formulation, when an agent uses AI, assisted performance is effectively anchored to the AI reliability term. In simplified form, the assisted-output rule can be written as
\[
Q_{HA}^{\mathrm{assist}} \approx Q_A.
\]
At the population level, hybrid performance is therefore governed by the mixture of autonomous and AI-assisted behavior, but the AI-assisted component itself does not include an explicit productive interaction between $Q_H$ and $Q_A$. Consequently, parameter tuning can improve retention or reduce dependency, but it cannot by itself create a mechanism for synergy.

To verify this interpretation, systematic sweeps were run across atrophy rate, AI reliability, AI adoption rate, dependency build rate, and dependency recovery rate. Across 2,640 parameter combinations in the baseline formulation, $CAI^*$ remained strictly negative, with a consistent ceiling near $-0.068$. Diagnostic sweeps using negative atrophy values, corresponding to AI-induced cognitive growth rather than erosion, did not remove this ceiling. This indicates that the constraint is structural rather than merely a consequence of unfavorable atrophy settings.

\begin{figure}[ht]
\centering
\IfFileExists{fig1_ceiling.pdf}{%
\includegraphics[width=0.85\linewidth]{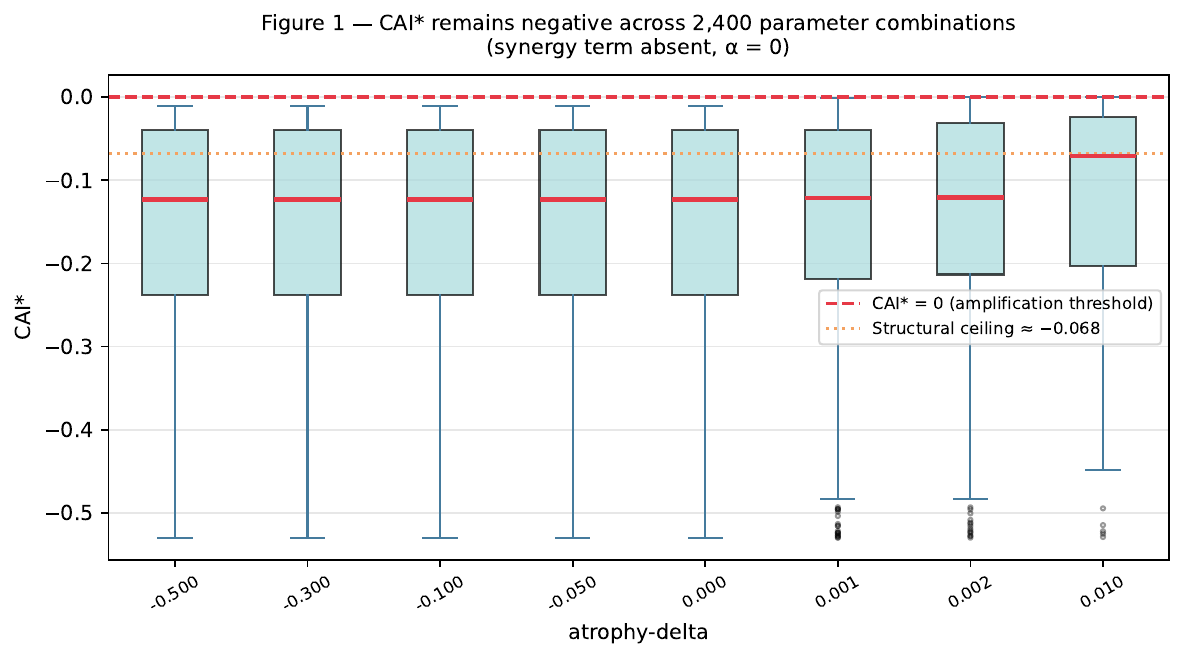}%
}{%
\fbox{\begin{minipage}{0.82\linewidth}
\centering
Placeholder for \texttt{fig1\_ceiling.pdf}.\\
Baseline parameter sweeps show a structural ceiling below $CAI^*=0$ when no explicit human--AI synergy term is present.
\end{minipage}}%
}
\caption{Distribution of $CAI^*$ across 2,640 parameter combinations in the baseline simulation. The structural ceiling near $-0.068$ persists regardless of atrophy rate, including diagnostic negative values. The amplification threshold is $CAI^*=0$.}
\label{fig:ceiling}
\end{figure}

\subsection{Introducing an explicit interaction term}

The conceptual model in Equation~\eqref{eq:synergy} includes a positive interaction term, $\alpha Q_H Q_A$, representing productive coupling between retained human capability and AI assistance. This term was not present in the baseline simulation rule. We therefore introduced a structural extension in which the assisted-performance computation becomes
\[
Q_{HA}^{\mathrm{assist}} = Q_A + \alpha Q_H Q_A,
\]
where $\alpha$ (implemented as \texttt{synergy-alpha}) controls the strength of the human--AI interaction. When $\alpha=0$, the assisted-output rule reduces to the baseline case. When $\alpha>0$, retained human capability is explicitly allowed to contribute to the hybrid output.

This modification should be interpreted as a structural test rather than as a literal empirical law of human--AI cognition. Its purpose is to determine whether positive $CAI^*$ requires an explicit mechanism by which human capability remains active in the assisted outcome, rather than being replaced by AI output.

\subsection{Results of the structural extension}

We evaluated 1,215 parameter combinations varying $\alpha \in [0,2]$, AI reliability in $\{0.5,0.7,0.9\}$, and AI adoption rate across the three experimental modes. Under this extended formulation, positive cognitive amplification became attainable. Specifically, $CAI^*>0$ was observed in 556 of 1,215 runs (46\%), with a maximum value of $0.504$. The amplification threshold appeared at approximately $\alpha \geq 0.25$.

\begin{figure}[ht]
\centering
\IfFileExists{fig2_synergy.pdf}{%
\includegraphics[width=\linewidth]{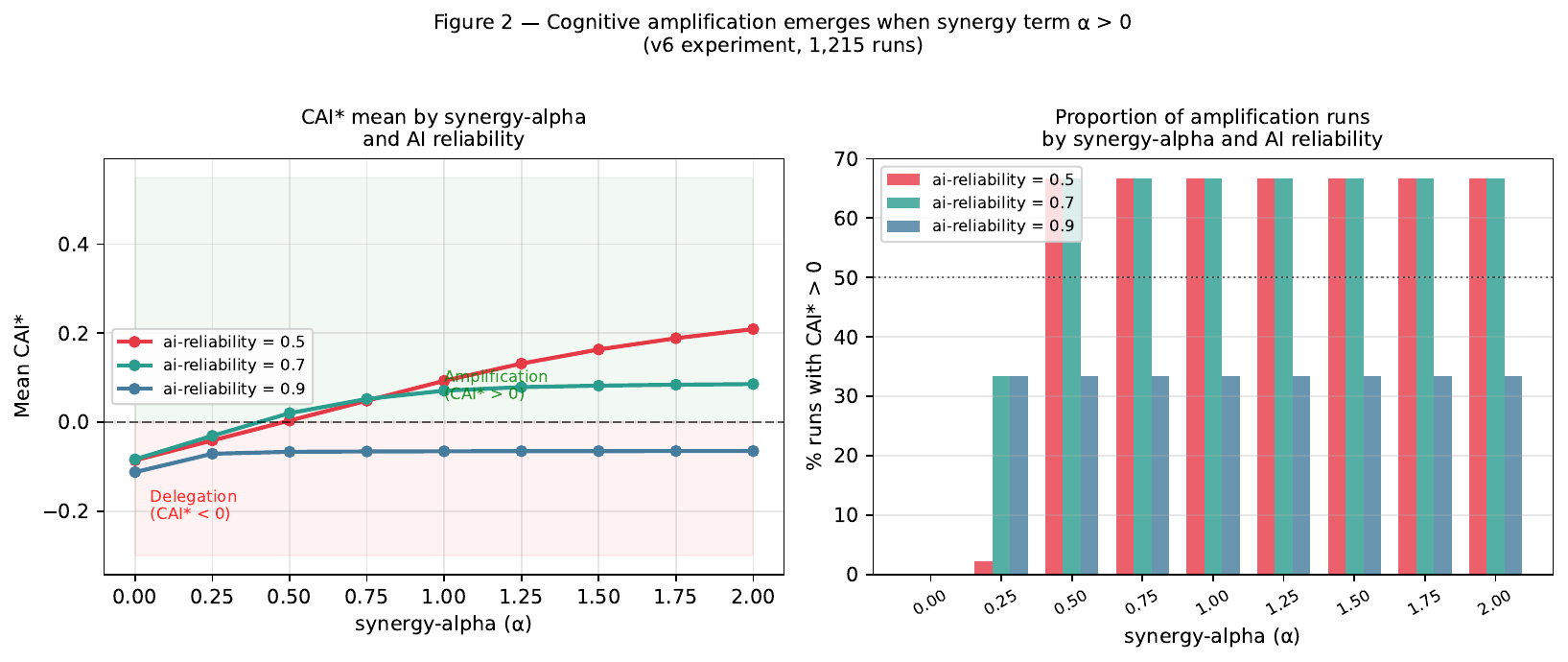}%
}{%
\fbox{\begin{minipage}{0.9\linewidth}
\centering
Placeholder for \texttt{fig2\_synergy.pdf}.\\
The structural-extension experiment shows mean $CAI^*$ and the proportion of runs achieving $CAI^*>0$ as functions of $\alpha$ and AI reliability.
\end{minipage}}%
}
\caption{Left: mean $CAI^*$ as a function of $\alpha$, stratified by AI reliability. Right: proportion of runs achieving $CAI^*>0$ by $\alpha$ and AI reliability. Cognitive amplification emerges consistently for $\alpha \geq 0.25$ when AI reliability is moderate ($0.5$--$0.7$), but does not emerge under high AI reliability ($0.9$) in the tested grid.}
\label{fig:synergy}
\end{figure}

A notable finding is that amplification did not emerge under high AI reliability ($Q_A=0.9$), but appeared consistently under moderate reliability ($Q_A=0.5$--$0.7$). This does not mean that weaker AI is intrinsically preferable. Rather, because $CAI^*$ is measured relative to $\max(Q_H,Q_A)$, a very strong standalone AI creates a demanding baseline that the hybrid system must exceed. Genuine amplification therefore requires not only high output quality, but also a coupling architecture in which human and AI capabilities combine productively enough to outperform the strongest standalone component.

\subsection{Implications for the metric framework}

These results refine the interpretation of the baseline experiments in three ways. First, the structural ceiling observed under $\alpha=0$ shows that $CAI^*>0$ is not achievable by parameter tuning alone when assisted performance collapses into AI output. Second, the emergence of amplification after introducing $\alpha Q_H Q_A$ shows that the framework can identify not only delegation and compromise regimes, but also the structural conditions under which genuine amplification becomes possible. Third, the results confirm that capability preservation and collaborative gain are distinct requirements: preventing atrophy may preserve $Q_H$, but positive amplification requires an interaction mechanism through which retained human capability contributes measurably to $Q_{HA}$.

Human--AI system designs that reduce human contribution to passive acceptance of AI outputs are therefore structurally constrained toward delegation-oriented regimes, even when they preserve some human capability. By contrast, designs that keep the human cognitively active may create the conditions under which genuine amplification can emerge.

\section{Discussion}

The framework proposed here does not attempt to measure intelligence in an absolute sense. Rather, it provides a relative metric system for evaluating whether human--AI interaction preserves, amplifies, or progressively displaces human cognitive capability.

The main contribution of the simulation is not merely to recover the familiar risk that strong short-term hybrid performance may coexist with long-term erosion of human capability. More importantly, it shows that a regime may remain apparently stable, preserve substantially more human capability than full delegation, and still fail to qualify as genuine amplification. In the experiments reported here, mixed reliance repeatedly occupies precisely this boundary position: it avoids outright collapse, approaches the AI baseline in hybrid performance, and yet remains structurally dependent, with negative collaborative gain and dependency above the balanced threshold.

This point matters because many practical evaluations of AI-assisted systems stop too early. A system that appears effective in terms of short-term output may still be structurally dependent on the artificial component, and may therefore fall short of true amplification even when the human agent has not yet visibly collapsed. The relevant distinction is therefore not simply between ``working'' and ``failing'' systems, but between systems that genuinely improve the joint cognitive unit and systems that merely maintain acceptable output under conditions of dependence.

This distinction is especially relevant in safety-critical and knowledge-intensive environments such as railway systems, aviation, pharmaceutical manufacturing, and medical technology. In such domains, the goal is not merely to achieve high nominal output, but to preserve robust human expertise under uncertainty, perturbation, and changing operational conditions. Excessive reliance on automation may therefore introduce a hidden form of fragility: systems can remain operationally efficient while progressively reducing the autonomy, adaptability, and resilience of the human component \cite{automation-bias,overreliance-ai}.

The experiments further suggest that cognitive delegation is not reducible to a single mechanism. It may arise through different combinations of dependency reinforcement and capability erosion. In some configurations, stronger dependency feedback pushes mixed reliance closer to effective delegation. In others, weaker behavioral feedback does not suffice to produce amplification when atrophy remains active. The optimization experiment strengthens this point: even under P2, the most favorable baseline configuration considered here, and even in the limiting case of zero atrophy, the system still fails to reach positive collaborative gain. This indicates that capability preservation matters, but is not by itself sufficient to recover genuine amplification under the present interaction dynamics.

The additional parameter sweeps and structural extension clarify why this failure occurs in the baseline model. If assisted performance is structurally anchored to the AI baseline, then parameter tuning can improve retention or reduce dependency, but it cannot create a productive interaction between retained human capability and AI support. Under that formulation, the absence of positive $CAI^*$ is not a weakness of the metric framework; it is a diagnosis of the interaction architecture.

Introducing the explicit synergy term changes this structural condition. Once the assisted output includes a contribution proportional to both retained human capability and AI performance, positive amplification becomes attainable in the simulated environment. This result strengthens the central claim of the paper: cognitive amplification is not simply a matter of using more reliable AI or preventing atrophy. It requires an interaction architecture in which human cognition remains an active and measurable contributor to the joint outcome.

From a design and evaluation standpoint, these results suggest a stronger claim than the original automation-trap formulation alone. Human--AI systems should not be judged solely by hybrid output, nor even by the absence of visible human collapse. They should instead be assessed through a joint lens that includes collaborative gain, dependency structure, retained human capability, and cognitive drift. Formally, the principle remains

\[
\max Q_{HA}
\quad \text{subject to} \quad
HCDR \ge 0,
\]

but the present results suggest that this condition is necessary rather than sufficient. A non-negative drift rate alone does not guarantee genuine amplification if the system remains structurally AI-dominated and fails to exceed the best standalone baseline. In other words, long-term preservation and collaborative gain must both be considered.

The broader implication is that genuine cognitive amplification may be substantially harder to obtain than conventional performance metrics suggest. This does not weaken the proposed framework; rather, it illustrates why such a framework is needed. Without metrics that jointly track collaborative gain, dependency, and retained human capability, one may easily mistake operational convenience for real augmentation.

The present work should therefore be understood not as a final theory of human--AI cognition, but as a decision-relevant evaluative lens. Its practical value lies in making visible whether a system is moving toward amplification, compromise, or delegation, and in showing that avoiding collapse is not the same as achieving genuine human--AI synergy.

\subsection{Limitations}
\subsubsection{Scope of the simulation.}
The agent-based simulation abstracts mechanisms of reliance and capability change 
rather than attempting to model the full phenomenology of human reasoning. Each 
agent's skill state is a compact numerical proxy for cognitive proficiency, and 
the update rules governing learning and atrophy are intentionally stylized. No 
claim is made that these rules reproduce the psychological or neurological 
processes underlying skill acquisition in human subjects. What the simulation 
does provide is a controlled environment in which the joint behavior of 
collaborative gain, dependency structure, and autonomous capability drift can be 
observed under systematically varied conditions. This is sufficient for the 
stated purpose: testing whether the metric framework produces informative and 
qualitatively distinct classifications across interaction regimes. Validation 
against longitudinal human-subject data---which would require extended, 
ethically supervised studies under controlled AI-assistance protocols---remains 
an important direction for future work.

The present results should therefore be read together with the empirical 
literature motivating the framework. Prior work provides evidence that 
overreliance, automation bias, cognitive offloading, and bias inheritance can 
affect human judgment and subsequent unaided reasoning. The contribution of the 
present paper is not to re-establish those phenomena empirically, but to provide a 
metric structure for evaluating whether such effects amount, at the system level, 
to amplification, compromise, or delegation.

\section{Future Work}

The present paper is primarily diagnostic rather than fully prescriptive. Its main contribution is to provide a compact metric framework for distinguishing among human-preserving, compromise, and delegation-dominated regimes of human--AI interaction. A central lesson of the current computational study is that genuine amplification appears difficult to obtain even in comparatively favorable simulated conditions. Several natural extensions therefore follow from this foundation.

A first direction is technical and remains close to the scope of the current work. The structural-extension experiment suggests that future optimization should vary not only atrophy and reliance parameters, but also the interaction architecture itself, including explicit synergy, explanation, scaffolding, and human-first mechanisms. More broadly, the proposed framework can be recast as the basis of a constrained optimization problem, in which a small number of design parameters---for example baseline AI reliance, dependency reinforcement, capability-preserving interventions, or interaction policies that affect learning under assistance---are adjusted so as to maximize collaborative gain while preserving human capability over time. In that setting, the relevant objective would not simply be to maximize hybrid output, but to search for regimes satisfying conditions such as $CAI^*>0$, $HCDR \ge 0$, and bounded dependency. This would allow the framework to move from regime identification toward principled search for viable amplification regions in parameter space.

A second direction concerns empirical grounding. The simulation developed here serves as a controlled stress test of the framework, but it remains an abstraction. Future work should therefore evaluate the proposed quantities in real human--AI interaction settings, particularly in domains where expertise retention, adaptation, and accountability are critical. Examples include industrial diagnosis, medical decision support, software engineering, and other environments in which AI assistance may improve immediate output while subtly altering long-term human competence.

A third direction concerns richer interaction design. In the present model, AI use is represented through stylized learning asymmetries, atrophy, and dependency feedback. Future work could examine more structured assistance policies, including delayed answer revelation, scaffolded hints, mandatory human-first attempts, periodic AI-off practice, or explanation-oriented assistance. Such mechanisms could be interpreted within the present framework as interventions that modify effective atrophy, dependency growth, or AI-mediated learning, thereby allowing systematic study of which design choices support genuine amplification rather than passive substitution.

Finally, broader normative and sociotechnical questions remain open. The framework introduced here can identify whether a system is moving toward amplification, compromise, or delegation, but it does not by itself determine what levels of autonomy, dependence, or expertise preservation ought to be considered acceptable in different institutional settings. Those questions extend beyond the present paper and may benefit from future work at the intersection of sociology, philosophy, human--computer interaction, education, and organizational studies. In that sense, the current work should be understood as an evaluative starting point: it offers a way to detect and compare cognitive regimes, while leaving the full normative and sociotechnical design problem for future research.

\section{Conclusion}

Artificial intelligence presents a dual possibility: it can strengthen human cognitive performance, or it can progressively displace the human role while preserving only the appearance of effective collaboration.

The metric framework proposed in this paper provides a way to analyze and monitor this distinction. Through a compact set of quantities---collaborative gain, dependency structure, human reliance, and cognitive drift---the framework makes visible differences that conventional performance evaluation can easily obscure. In particular, the agent-based experiments show that strong hybrid performance does not by itself imply genuine cognitive amplification. A regime may preserve more human capability than full delegation and still remain structurally dependent on the artificial component, failing to exceed the best standalone baseline.

The baseline simulation results distinguish three practically relevant regimes: degenerate AI-dominated delegation, human-preserving but weakly competitive interaction, and intermediate boundary regimes that approach the AI baseline while still failing to achieve positive collaborative gain. The optimization experiment strengthens this conclusion: even under the most favorable baseline configuration considered here, and even when atrophy is reduced to zero, genuine amplification is not recovered. This suggests that preventing capability erosion is necessary for cognitive sustainability, but not sufficient for amplification.

The additional parameter sweeps and structural-extension experiment sharpen this conclusion. Within the baseline interaction architecture, amplification remains unattainable by parameter tuning alone. Positive $CAI^*$ becomes attainable only after introducing an explicit mechanism through which retained human capability contributes to the assisted output. This indicates that genuine amplification is not merely a consequence of stronger AI, greater adoption, or lower atrophy. It is a property of the coupling architecture between human and AI.

The broader implication is that evaluating human--AI systems requires more than measuring immediate hybrid output. Systems should also be assessed in terms of whether they preserve human capability, avoid structural dependence, and support cognitively sustainable collaboration over time. In safety-critical and knowledge-intensive domains, the goal of artificial intelligence should not be the silent replacement of human cognition, but the preservation and strengthening of expert judgment.

Systems that improve output while degrading retained human capability may increase operational convenience, but they do not necessarily increase the intelligence of the overall human--AI system. Genuine amplification requires both sustained human competence and an interaction design that allows human and artificial capabilities to combine productively.

\appendix

\section{Optimization procedure}
\label{app:optimization}

To test whether capability-preserving intervention alone can produce genuine cognitive amplification, we performed a constrained parameter search over the atrophy rate $\delta$ (implemented as \texttt{atrophy-delta}) using NetLogo BehaviorSpace.

\subsection{Search space}

The search was restricted to the mixed-reliance regime, with all simulation parameters held fixed except $\delta$. We evaluated a predefined grid of candidate values:
    \[
\delta \in \{0.0040,\,0.0035,\,0.0030,\,0.0025,\,0.0020,\,0.0015,\,0.0010,\,0.0005,\,0.0000\}.
\]

Each candidate value was evaluated over 20 random seeds, using the same simulation horizon and final aggregation window as in the main experiments.

\subsection{Optimization criterion}

The search objective was defined in terms of the framework proposed in the main text. For each candidate value of $\delta$, we computed the mean values of $CAI^*$, $D$, and $HCDR$ across seeds. We then applied the following constrained selection rule:

\[
\delta^\star
=
\arg\max_{\delta}
\; CAI^*(\delta)
\quad
\text{subject to}
\quad
HCDR(\delta) \ge 0.
\]

Thus, the procedure seeks the value of $\delta$ that maximizes collaborative gain while excluding solutions that degrade human capability over time.

\subsection{Decision rule}

A candidate configuration was considered to achieve genuine amplification only if it satisfied both:
\[
CAI^* > 0
\qquad \text{and} \qquad
HCDR \ge 0.
\]

In the event that no candidate satisfies these conditions, the search still identifies the best attainable compromise under the imposed constraint. In that case, the optimization result is interpreted as evidence that reducing atrophy alone is insufficient to produce genuine amplification under the current interaction dynamics.

\subsection{Implementation}

The search was implemented directly in BehaviorSpace as a one-parameter sweep over \texttt{atrophy-delta}. For each value of $\delta$, the model was run over the full simulation horizon and the reported metrics were aggregated over the final evaluation window. This yields a reproducible and transparent constrained search procedure aligned with the metric framework introduced in the paper.

\subsection{Full results of the atrophy-parameter search}
\label{app:atrophy-full-results}
Table~\ref{tab:atrophy-grid-search} reports the full constrained search over the atrophy parameter $\delta$ under mixed reliance.
\begin{table}[htbp]
\centering
\resizebox{\textwidth}{!}{%
\begin{tabular}{
@{}
S[table-format=1.4]
S[table-format=-1.4]
S[table-format=1.4]
S[table-format=1.4]
S[table-format=1.4]
S[table-format=1.4]
S[table-format=1.4]
S[table-format=1.4]
S[table-format=1.4]
S[table-format=1.4]
S[table-format=1.4]
S[table-format=1.4]
S[table-format=1.4]
S[table-format=1.4]
S[table-format=1.4]
@{}
}
\toprule
& \multicolumn{2}{c}{$CAI^*$}
& \multicolumn{2}{c}{$D$}
& \multicolumn{2}{c}{$HCDR$}
& \multicolumn{2}{c}{$Q_H$}
& \multicolumn{2}{c}{Skill}
& \multicolumn{2}{c}{$Q_{HA}$}
& \multicolumn{2}{c}{AI use} \\
\cmidrule(lr){2-3}
\cmidrule(lr){4-5}
\cmidrule(lr){6-7}
\cmidrule(lr){8-9}
\cmidrule(lr){10-11}
\cmidrule(lr){12-13}
\cmidrule(lr){14-15}
{$\delta$}
& {Mean} & {Std}
& {Mean} & {Std}
& {Mean} & {Std}
& {Mean} & {Std}
& {Mean} & {Std}
& {Mean} & {Std}
& {Mean} & {Std} \\
\midrule
0.0040 & -0.0351 & 0.0032 & 1.0374 & 0.0034 & 0.0000 & 0.0037 & 0.6059 & 0.0269 & 0.3279 & 0.0086 & 0.9649 & 0.0032 & 0.8361 & 0.0153 \\
0.0035 & -0.0343 & 0.0031 & 1.0364 & 0.0033 & 0.0000 & 0.0037 & 0.6112 & 0.0270 & 0.3535 & 0.0096 & 0.9657 & 0.0031 & 0.8289 & 0.0157 \\
0.0030 & -0.0333 & 0.0029 & 1.0354 & 0.0031 & 0.0000 & 0.0037 & 0.6184 & 0.0265 & 0.3892 & 0.0115 & 0.9667 & 0.0029 & 0.8221 & 0.0166 \\
0.0025 & -0.0309 & 0.0028 & 1.0327 & 0.0029 & 0.0001 & 0.0037 & 0.6324 & 0.0266 & 0.4498 & 0.0131 & 0.9691 & 0.0028 & 0.8076 & 0.0184 \\
0.0020 & -0.0276 & 0.0031 & 1.0291 & 0.0033 & 0.0001 & 0.0037 & 0.6506 & 0.0267 & 0.5423 & 0.0142 & 0.9724 & 0.0031 & 0.7899 & 0.0182 \\
0.0015 & -0.0237 & 0.0032 & 1.0250 & 0.0034 & 0.0001 & 0.0037 & 0.6761 & 0.0254 & 0.6694 & 0.0117 & 0.9763 & 0.0032 & 0.7733 & 0.0189 \\
0.0010 & -0.0219 & 0.0033 & 1.0231 & 0.0034 & 0.0001 & 0.0037 & 0.6958 & 0.0249 & 0.7917 & 0.0068 & 0.9781 & 0.0033 & 0.7670 & 0.0192 \\
0.0005 & -0.0209 & 0.0033 & 1.0221 & 0.0035 & 0.0001 & 0.0037 & 0.7055 & 0.0236 & 0.8878 & 0.0039 & 0.9791 & 0.0033 & 0.7636 & 0.0194 \\
0.0000 & -0.0207 & 0.0032 & 1.0218 & 0.0034 & 0.0001 & 0.0037 & 0.7125 & 0.0227 & 1.0000 & 0.0000 & 0.9793 & 0.0032 & 0.7630 & 0.0194 \\
\bottomrule
\end{tabular}%
}
\caption{Constrained search over the atrophy parameter \texorpdfstring{$\delta$}{delta} under mixed reliance. Values are mean and standard deviation over 20 random seeds. Lower atrophy improves retained human capability and collaborative gain, but even zero atrophy does not yield positive collaborative gain.}
\label{tab:atrophy-grid-search}
\end{table}

\section*{Declarations}

\subsection*{Author Contribution}
Eduardo Di Santi conceived the study, developed the theoretical and metric framework, designed the baseline simulation architecture, performed the baseline analysis and atrophy-optimization study, interpreted the results, and wrote and revised the manuscript. Carla Florida made a substantial contribution to the empirical and computational development of the paper by conducting additional parameter sweeps, testing whether amplification could emerge within the original parameter space, identifying the structural ceiling of the baseline model, implementing the structural extension with an explicit human--AI synergy term, and generating and analyzing the corresponding figures and simulation evidence. Carla Florida also contributed text for the structural-extension section. Both authors discussed the interpretation of the results, reviewed the manuscript, and approved the submitted version.

\subsection*{Availability of Data and Materials}
The computational code supporting the findings of this study, including the NetLogo simulation and associated experimental materials required to reproduce the reported experiments, is available at:
\url{https://anonymous.4open.science/status/cognitive_amplification_vs_delegation-25A9}

\bibliographystyle{unsrt}
\bibliography{sn-bibliography}

\end{document}